\documentclass{JHEP}
\usepackage{epsfig}

\makeatletter
\def\setlabel#1{\gdef\@eqnnum{{#1}}}
\def\restlabels{\gdef\@eqnnum{{\normalfont \normalcolor (\theequation)}}}
\makeatother

\title{Day-Night and Energy Dependence of MSW Solar Neutrinos for Maximal Mixing}
\author{Alan H.~Guth,  Lisa Randall,  Mario Serna \\  guth@ctp.mit.edu, randall@baxter.mit.edu, mariojr@mit.edu
\\  Center for Theoretical Physics \\ Massachusetts Institute of Technology\\ Cambridge, Massachusetts 02139, U.S.A.\\ }
\abstract{It has been stated in the literature that the case of maximal
mixing angle for $\nu_e$ leads to no day-night effect for solar
neutrinos and an energy independent flux suppression of
$\frac{1}{2}$.  While the case of maximal mixing angle and
$\Delta m^2$ in the MSW range does lead to suppression of the
electron neutrinos reaching the earth from the sun by $P_S={1
\over 2}$, the situation is different for neutrinos that have
passed through the earth.  We make the pedagogical point that,
just as with smaller mixing angles, the earth regenerates the
$|\nu_1 \rangle$ state from the predominantly $|\nu_2 \rangle$
state reaching the earth, leading to coherent interference
effects.  This regeneration can lead to a day-night effect and an
energy dependence of the suppression of solar electron neutrinos,
even for the case of maximal mixing.  For large mixing angles,
the energy dependence of the day-night asymmetry depends heavily
on $\Delta m^2$. With a sufficiently sensitive measurement of the
day-night effect, this energy dependence could be used to
distinguish among the large mixing angle solutions of the solar
neutrino problem.}
\preprint{DISTRIBUTION A. \\ hep-ph/9903464 \\ MIT-CTP-2841 \\ Revised April 28, 1999\\}
\begin{document}

\section{Introduction}

The recent Super-Kamiokande announcement that atmospheric
neutrinos are nearly maximally mixed has renewed much interest in
the possibility that solar neutrinos might also be maximally
mixed.  In this paper we will consider only two-neutrino mixings,
so by ``maximal mixing'' we are referring to the possibility that
the two lightest mass eigenstates, $|\nu_1
\rangle$ and $|\nu_2 \rangle$, with eigenvalues $m_1$ and $m_2$
respectively ($m_1 < m_2$), are each equal-probability
superpositions of the flavor eigenstate $|\nu_e \rangle$
(electron neutrino) and some other state $|\nu_x \rangle$, where
$|\nu_x \rangle$ can be any linear combination of $|\nu_\mu
\rangle$ (muon neutrino) and $|\nu_\tau \rangle$ (tau neutrino).
Many theoretical models have been proposed predicting the
possibility of such maximal mixing (for example, see
\cite{Xing1,hepph9806387,hepph9806540,hepph9807325,Xing2,hepph9808296,hepph9811378}).
In this paper we are concerned only with the MSW solutions to the
solar neutrino problem, first proposed by Mikheyev, Smirnov, and
Wolfenstein \cite{MSW1,MSW2,MSW3}, while the alternative
possibility of nearly maximally mixed vacuum oscillations has
been considered by other authors \cite{hepph9903262}.  The MSW
effect results from the neutrino interaction with matter, causing
an enhancement of the conversion process transforming $\nu_e$
into $\nu_x$. The MSW effect is also capable of driving neutrinos
back towards a $\nu_e$ state after passing through the earth.
This process would result in a change in the $\nu_e$ flux between
daytime and nighttime measurements, a phenomenon known as the
day-night effect , or more generally zenith angle dependence.
Over the past decade there have been extensive studies of the
day-night effect
\cite{Carlson,BaltzWeneser,Baltz2,hepph9702343,hepph9702361,hepph9705392,hepph9706239,hepex9812009},
which have been mostly concerned with the small mixing angle
solutions to the solar neutrino problem.  Most of these studies
have used the Mikheyev-Smirnov expression \cite{MS2} to describe
the effect of the earth on the solar neutrinos, which we will
hereafter refer to as Eq.~(\ref{MainEq}):
\begin{equation}
P_{SE} = \frac{P_S - \sin^2 \theta_V + P_{2e}(1-2 P_S)}{\cos2
\theta_V} \ .
\label{MainEq}
\end{equation}
Here $P_S$ is the probability that an electron neutrino ($| \nu_e
\rangle$) originating in the sun will be measured as an electron
neutrino upon reaching the earth, $P_{SE}$ is the probability
that an electron neutrino originating in the sun will be measured
as an electron neutrino after passing through the earth, $P_{2e}$
is the probability that a pure $| \nu_2 \rangle$ eigenstate
entering the earth will be measured as an electron neutrino when
it emerges, and $\theta_V$ is the vacuum mixing angle, defined by

\begin{eqnarray}
\label{ThetaDef}
|\nu_1 \rangle & = & |\nu_e \rangle \cos \theta_V - |\nu_x \rangle
\sin \theta_V \ ,\setlabel{(1.2a)} \\
|\nu_2 \rangle & = & |\nu_x \rangle \cos \theta_V + |\nu_e \rangle
\sin \theta_V \ .\setlabel{(1.2b)}
\end{eqnarray}
\restlabels
\addtocounter{equation}{-1}

In the previous studies of the day-night effect several authors
have claimed that there is no day-night effect at $P_S=1/2$
\cite{hepph9706239,Baltz2}.  In fact, in the formula for the
day-night effect which is conventionally used,
Eq.~(\ref{MainEq}), the properties of the earth enter only
through $P_{2e}$, which is explicitly multiplied by $(1-2 P_S)$. 
We wish to emphasize, however, that the case of maximal mixing is
an exception to this statement.  For maximal mixing
Eq.~(\ref{MainEq}) is ill-defined, since $\cos 2\theta_V = 0$,
and we will show below that generically there is a day-night
effect for this case.  Nonetheless, we have no disagreements with
either the equations or the contour plots in the aforementioned
papers, which in fact do show non-zero day-night effects at
maximal mixing.  The purpose of this paper is to clarify the
previous papers, and also to investigate more carefully the role
of the day-night effect for maximal mixing. We will show that at
maximal mixing $P_{SE} \neq 1/2$, implying a day-night effect and
an often overlooked energy-dependence of the suppression of the
solar neutrino flux.

Physically, the day-night effect survives because the neutrino
beam reaching the earth, for all MSW solutions, is predominantly
$| \nu_2 \rangle$.  For maximal mixing this state is half $\nu_e$
and half $\nu_x$, but there is a definite phase relationship, $|
\nu_2 \rangle = (| \nu_e \rangle + | \nu_x \rangle)/\sqrt{2}$,
so the density matrix is not proportional to the identity matrix. 
A coherent component of $| \nu_1 \rangle$ is regenerated as this
beam traverses the earth, leading to interference with the
incident $| \nu_2 \rangle$ beam.  The case is rather different
from the small mixing-angle case, for which Eq.~(\ref{MainEq})
really does imply the absence of a day-night effect when $P_S =
1/2$.  For a small mixing angle $P_S$ equals $1/2$ only when
conditions in the sun drive the ensemble into a density matrix
proportional to the identity matrix, in which case the earth
would have no effect.

In the remainder of this paper we explain in more detail why
maximal mixing can result in a day-night effect.  In Section
\ref{derivation}, we review the derivation of Eq.~(\ref{MainEq})
as given by Mikheyev and Smirnov \cite{MS2}, and we resolve the
maximal mixing ambiguity.  Next in Section \ref{numericalresults}
we present results of numerical calculations showing the
day-night effect at maximal mixing. Finally in the appendices, we
provide greater details concerning the analytic and numerical
calculations presented in the paper.

\section{Derivation of Equation (1.1) }
\label{derivation}

The key assumption necessary for the derivation of
Eq.~(\ref{MainEq}) is that the neutrino beam arriving at the
earth can be treated as an incoherent mixture of the two mass
eigenstates $| \nu_1 \rangle$ and $| \nu_2 \rangle $.  That is,
we assume that there is no interference between the $\nu_1$ and
$\nu_2$ components reaching the earth, or equivalently that the
off-diagonal entries of the density matrix in the $\nu_1$-$\nu_2$
basis are negligibly small.  The physical effects which cause
this incoherence are discussed in Appendix \ref{DeCoherance}. In
the case of maximal mixing, the incoherence is ensured for
$\Delta m^2 > 6.5 \times 10^{-9} \ {\rm eV}^2$ because of the
energy resolution of current detectors.  Other sources of
incoherence include the separation of $| \nu_1 \rangle$ and $|
\nu_2 \rangle$ wave packets in transit to the earth, the
averaging over the regions in the sun where the neutrinos were
produced, and the averaging over the changing radius of the
earth's orbit \cite{hepph9903329}.  In Appendix \ref{DeCoherance}
we comment on
     the regions of parameter space
     for which the assumption of incoherence is
     valid.

Given the assumption of incoherence, we write the fractions of $|
\nu_1 \rangle$ and $| \nu_2 \rangle$ flux from the sun as $k_1$
and $k_2$, respectively\footnote{For large mixing angles, $\sin^2
2 \theta_V \geq 0.5$ and $5 \times 10^{-5} \leq \Delta m^2
(\rm{eV})^2 \leq 1 \times 10^{-7}$, $k_2 \approx 1$ and $k_1
\approx 0$.}.  Since there is no interference, the probability
that a solar neutrino will be measured as $\nu_e$ upon reaching
the surface of the earth is given by
\begin{eqnarray}
P_S & = & k_1 \left|\langle\nu_e|\nu_1\rangle\right|^2 + k_2
\left|\langle\nu_e|\nu_2\rangle\right|^2 \nonumber \\
  & = & k_1 \cos^2 \theta_V + k_2 \sin^2 \theta_V \nonumber \\
  & = & \cos^2 \theta_V - k_2 \cos 2 \theta_V \ ,
\label{PSeq}
\end{eqnarray}
where we have used Eqs.~(\ref{ThetaDef}) and the fact that $k_1 +
k_2 = 1$. Similarly, the probability that a solar neutrino will
be measured as $\nu_e$ after passing through the earth, when it
is no longer in an incoherent superposition of the mass
eigenstates, is given by
\begin{equation}
P_{SE} = k_1 P_{1e} + k_2 P_{2e} \ ,
\label{Psenext}
\end{equation}
where $P_{1e}$ ($P_{2e}$) is the probability that a $| \nu_1
\rangle$ ($|\nu_2 \rangle$) eigenstate will be measured as
$\nu_e$ after traversing the earth.  Finally, the unitarity of
the time evolution operator implies that the state vectors of two
neutrinos entering the earth as $|\nu_1 \rangle$ and $|\nu_2
\rangle$ must remain orthonormal as they evolve through the earth
and become $|\tilde \nu_1 \rangle$ and $|\tilde \nu_2 \rangle$,
respectively.  Therefore
\begin{equation}
P_{1e} + P_{2e} = \left|\langle\nu_e | \tilde
\nu_1\rangle\right|^2 +
\left|\langle\nu_e | \tilde \nu_2\rangle\right|^2  = 1 \ .
\label{Unitarity}
\end{equation}
Eq.~(\ref{MainEq}) can then be obtained by using Eq.~(\ref{PSeq})
and the above equation to eliminate $P_{1e}$, $k_1$, and $k_2$
from Eq.~(\ref{Psenext}).

From the above derivation, one can see that the singularity of
Eq.~(\ref{MainEq}) at maximal mixing arises when Eq.~(\ref{PSeq})
is solved to express $k_2$ in terms of $P_S$.  For maximal mixing
$P_S = 1/2$ for any value of $k_2$, so $k_2$ cannot be expressed
in terms of $P_S$.  The ambiguity disappears, however, if one
leaves $k_2$ in the answer, so Eq.~(\ref{Unitarity}) can be used
to rewrite Eq.~(\ref{Psenext}) as
\begin{equation}
P_{SE} = {1 \over 2} + 2 \left( k_2 - {1 \over 2} \right) \left(P_{2e} -
{1 \over 2} \right) \ .
\label{Psehalf}
\end{equation}
Thus, $P_{SE} = 1/2$ only if $k_2 = 1/2$ or $P_{2e} = 1/2$.  For
the MSW solutions at maximal mixing one has $k_2 \approx 1$, and
there is no reason to expect $P_{2e} = 1/2$.  Generically $P_{SE}
\neq 1/2$ for the case of maximal mixing.

\section{The day-night effect at maximal mixing}
\label{numericalresults}

Using the evolution equations derived in Appendix
\ref{MSWDerivation} and the procedures described in Appendix
\ref{CalcDetails}, we have calculated a variety of properties
concerning the day-night effect for maximal mixing angle. The
calculation parameters are chosen for those of the
Super-Kamiokande detector.

\begin{figure}
\centerline{\epsfig{figure=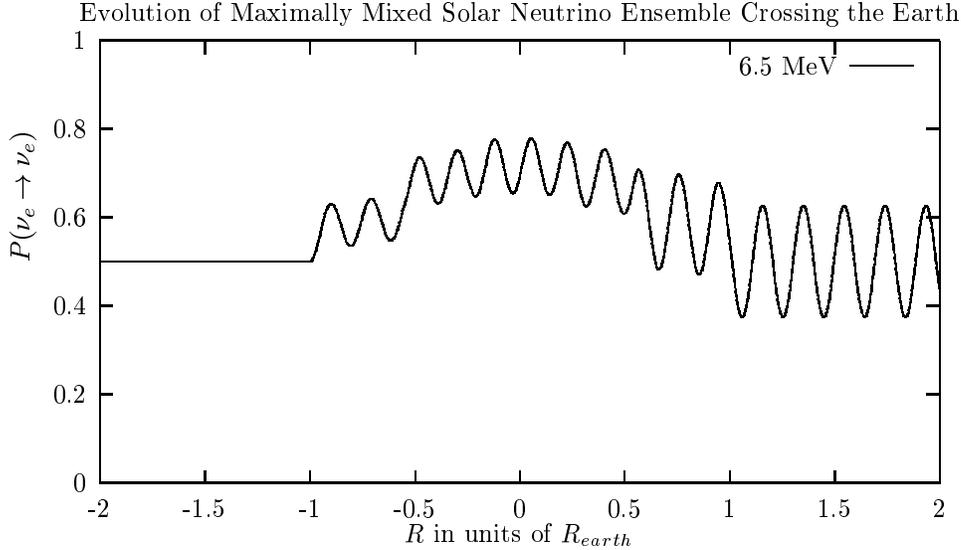,angle=0}}
\caption{The evolution of $P_{\nu_e \rightarrow \nu_e}$ as the
ensemble of neutrinos propagates across the center of the earth. 
The neutrinos enter the earth as an incoherent mixture of the
energy eigenstates $\nu_1$ and $\nu_2$ which is almost completely
$\nu_2$.  This plot shown is for $\Delta m^2 = 1.3 \times 10^{
-5} \ {\rm eV}^2$ and a neutrino energy $E = 6.5 \ {\rm MeV}$. }
\label{MMTimeSeries} 
\end{figure}

Figure \ref{MMTimeSeries} shows the evolution of $P(\nu_e
\rightarrow \nu_e)$, the probability that a solar neutrino will
be measured as $\nu_e$, as the beam of neutrinos traverses a path
through the center of the earth. Notice that after traversing the
earth the ensemble of neutrinos is no longer in a steady state,
but instead $P(\nu_e \rightarrow \nu_e)$ continues to oscillate
in the vacuum. From the perspective of the mass eigenstates, the
neutrinos under consideration arrive at the earth roughly in a $|
\nu_2 \rangle$ state.  Upon reaching the earth, the
step-function-like changes in the electron density profile (see
Fig.~\ref{EarthProfile}) cause non-adiabatic evolution,
regenerating the $| \nu_1 \rangle$ state and leading to
interference effects.  In the regions of parameter space where
the day-night effect is maximal because the oscillation length of
these interference terms coincides with the length of the slabs
of near constant density composing the earth, the resulting
buildup of $\nu_e$ flux has been called oscillation length
resonance \cite{hepph9805262,hepph9811205,hepph9903302}.

\begin{figure}
\centerline{\psfig{figure=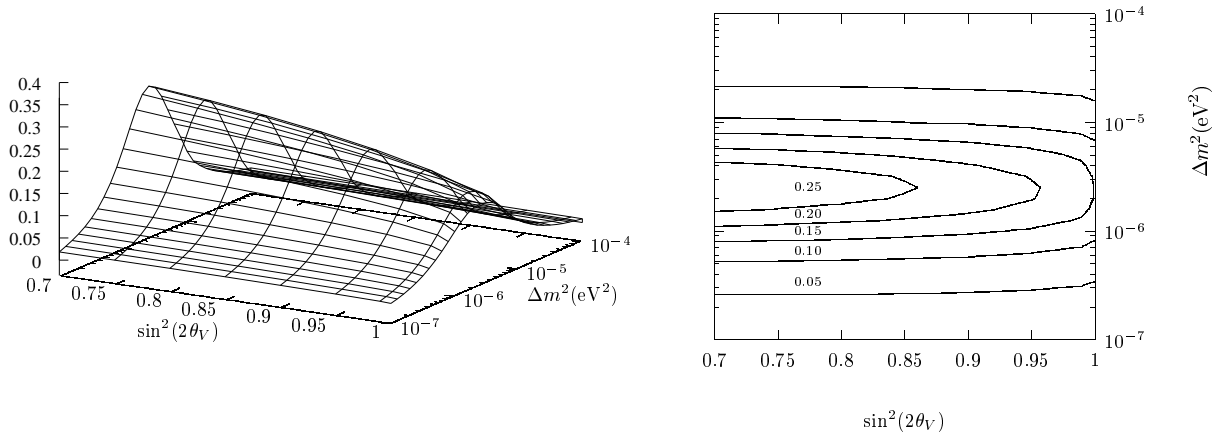,width=6.8in,angle=0}}
\caption{The day-night asymmetry ($A_{d-n}=(N-D)/(N+D)$) as a
function of mixing parameters calculated using the density
matrix.  On the left is a three dimensional surface where the
height of the surface is the day-night asymmetry.  Notice that
the exposed edge is calculated at maximal mixing and is clearly
non-zero. On the right is a contour plot showing the lines of
constant day-night asymmetry.  }
\label{ContourDM}
\end{figure}

In Fig.~\ref{ContourDM} we present a contour plot calculated from
the density matrix that exemplifies the non-zero nature of the
day-night effect at maximal mixing.  On the left is a
three-dimensional surface where the height of the surface is the
day-night asymmetry.  Notice that the exposed edge is calculated
at maximal mixing and is clearly non-zero. On the right is a
contour plot showing the lines of constant day-night asymmetry, a
plot which is identical to those produced in other references. 

\begin{figure}
\centerline{\psfig{figure=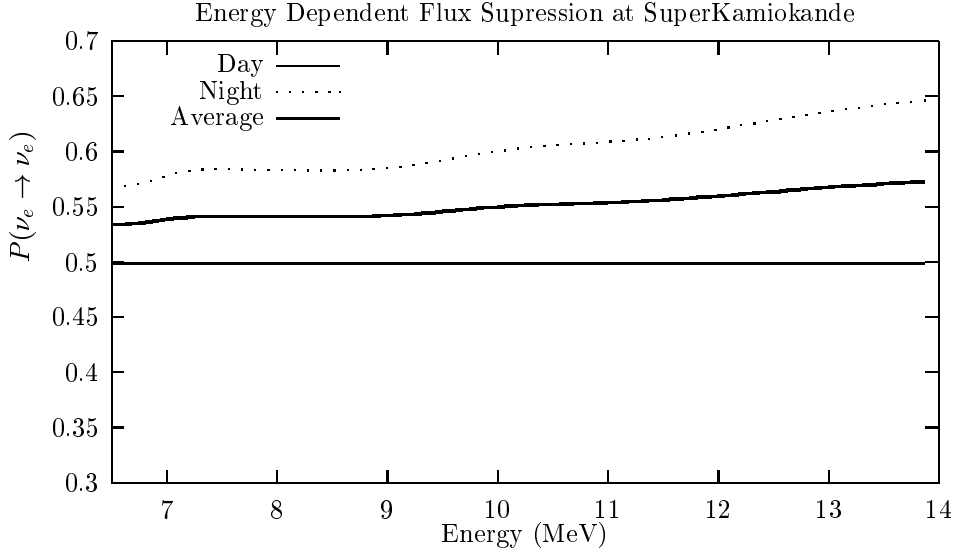,angle=0}}
\caption{The predicted flux suppression as a function of energy. 
Notice that the predicted overall flux suppression is not 1/2,
due to day-night effects, even though the mixing angle is
maximal.  The plot is for $\Delta m^2=1.0 \times 10^{-5} \ {\rm
eV}^2 $ which is near the border of the region excluded by the
small day-night effect ($A_{d-n}$) measured at Super-Kamiokande.}
\label{MMEnergy}
\end{figure}

We now explain how the inclusion of the day-night effect at
maximal mixing resolves a certain confusion that has arisen in
the past because of its neglect.  As a result of the non-zero
day-night effect, there exists an energy dependence at maximal
mixing, as can be seen in Fig.~\ref{MMEnergy}.  If one assumes
that the flux suppression at maximal mixing has no energy
dependence, as was done in Ref.~\cite{hepph9810272}, then there
is an apparent discrepancy between two sections of
Ref.~\cite{hepph9807216}. Sec.~IV-D excludes the possibility of
energy-independent oscillation into active (as opposed to
sterile) neutrinos at the 99.8\% confidence level, while Fig.~2
shows some regions of the maximal-mixing-angle parameter space
not excluded at the 99\% confidence level. 
Ref.~\cite{hepph9810272} has tried to resolve this discrepancy
without including the day-night effect, concluding that maximal
mixing is excluded at the 99.6\% confidence level.  The actual
resolution to this apparent discrepancy is that Fig.~2 of
Ref.~\cite{hepph9807216} includes the energy dependence induced
by the day-night effect at maximal mixing, while Sec.~IV-D
discusses the case of energy-independent flux suppression and
does not apply to maximal mixing.  The correct conclusion is that
of Fig.~2, which shows that maximal mixing is not excluded at the
99\% confidence level.

Whether or not the day-night effect is included, maximal mixing
is not a very good fit to the experimental data from the three
neutrino experiments (chlorine, gallium, and water)
   \cite{hepph9807216}.  However, maximal mixing does fit well if
the chlorine data is excluded on the suspicion of some systematic
error \cite{hepph9807235}.  Ref.~\cite{hepph9806540} has argued
that if the $^8B$ flux is about 17\% lower than the standard
solar model (BP98)
   \cite{BP98}, then a bi-maximal mixing scenario becomes a
tenable solution to the solar neutrino problem.  The MSW
mechanism described here is applicable for $\Delta m^2 > 6.5
\times 10^{-9}\ \rm{eV}^2$.  In the bi-maximal mixing scenario
that we consider the upper bound on $\Delta m^2$ is set by the
CHOOZ data constraining $\Delta m^2 \leq 0.9 \times 10^{-3}\
\rm{eV}^2$
\cite{hepex9711002}.

\begin{figure}
\centerline{\psfig{figure=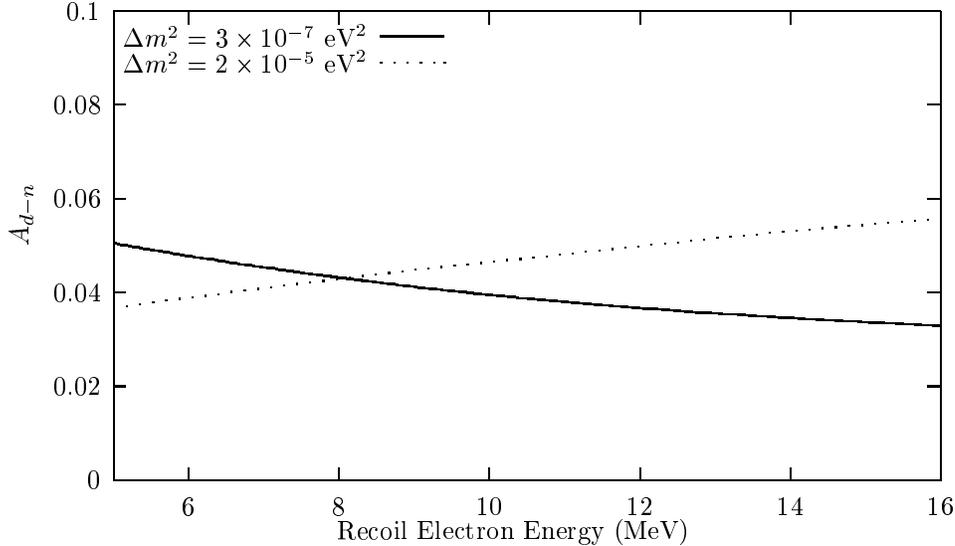,angle=0}}
\caption{The day-night asymmetry ($A_{d-n}=(N-D)/(N+D)$) as a
function of recoil electron energy at Super-Kamiokande.  Both
plots are at maximal mixing angle, with $\Delta m^2$ at the upper
and lower borders of the region disfavored by the smallness of
the day-night effect observed at Super-Kamiokande. The rising
line is for $\Delta m^2=2 \times 10^{-5}\ \rm{eV}^2$, and the
descending line is for $\Delta m^2=3 \times 10^{-7}\ \rm{eV}^2$.
}
\label{Adnvsnrg}
\end{figure}

When detailed studies of the day-night effect are completed, the
energy (and zenith angle) dependence will be valuable additional
information. To the best of our knowledge, the Super-Kamiokande
Collaboration has not published their day-night asymmetry as a
function of recoil electron energy. Past studies of the day-night
effect have noted the energy dependence of the day-night
asymmetry \cite{hepph9705392,hepph9706239}.  While for small
mixing angles $|A_{d-n}| < 0.02$ without a clear energy
dependence
\cite{hepph9706239}, for large mixing angles the $A_{d-n}$ energy
dependence can be significant and informative.
Fig.~\ref{Adnvsnrg} shows the theoretical predictions of the
day-night asymmetry in the electron recoil spectrum at
Super-Kamiokande for two cases of maximal mixing: $\Delta m^2 = 2
\times 10^{-5}\ \rm{eV}^2$ and $\Delta m^2=3 \times 10^{-7}\
\rm{eV}^2$.  Note that the two curves have opposite slopes.

The approximate shape of the graph of $A_{d-n}$ vs.~recoil
electron energy can be understood from Fig.~\ref{ContourDM},
using the fact that Fig.~\ref{ContourDM} is dominated by the peak
of the $^8B$ neutrino spectrum at about 6.5 \hbox{MeV}.  It is
shown in Appendix \ref{MSWDerivation} that the neutrino evolution
equations (Eqs.~(\ref{eom})-(\ref{BdefApp})) depend on $\Delta
m^2$ and the neutrino energy (or momentum) $E$ only through the
combination $\Delta m^2 / E$.  Thus, Fig.~\ref{ContourDM} shows
that for any value of $\sin^2 2 \theta_V$, $A_{d-n}$ has a
maximum at $\Delta m^2 / E \approx 2.5 \times 10^{-6} \ {\rm
eV}^2 / (6.5 \ {\rm MeV})$.  When $E$ is varied at fixed $\Delta
m^2$, $A_{d-n}$ will have a peak at
\begin{equation}
E \approx {\Delta m^2 \over 2.5 \times 10^{-6} \ {\rm eV}^2}
\times 6.5 \ {\rm MeV} \ .
\end{equation}
So for $\Delta m^2 = 2 \times 10^{-5}\ {\rm eV}^2$ the peak lies
far to the right of the scale in Fig.~\ref{Adnvsnrg}, so the
curve slopes upward.  For $\Delta m^2 = 3 \times 10^{-7}\ {\rm
eV}^2$ the peak lies far to the left, and the curve slopes
downward.

Fig.~\ref{ContourDM} shows that the peak in the graph of
$A_{d-n}$ vs.\ $\Delta m^2$ is higher at large mixing angles
($\sin^2 2 \theta_V \approx 0.7$) than it is at maximal mixing,
so the same will be true for the energy dependence of the
day-night effect. For $\sin^2 2 \theta_V=0.63$ and $\Delta
m^2=1.3 \times 10^{-5}\ \rm{eV}^2$, for example, the slope of the
graph of $A_{d-n}$ vs.\ recoil electron energy is about twice the
magnitude of the slopes shown in Fig.~\ref{Adnvsnrg}. Thus, the
day-night asymmetry as a function of recoil electron energy could
be a strong indicator of $\Delta m^2$ if the solar neutrinos have
a large or maximal mixing angle in the MSW range of parameters.

\section{Conclusions}

We have disproved the assumption that $P_S=1/2$ always implies
$P_{SE}=1/2$.  We have also shown that neutrinos with a maximal
mixing angle can have a day-night effect and that they do not
always result in a uniform energy-independent flux suppression of
$1/2$.  Because the issues that we have attempted to clarify
concern mainly the words that have been used to describe correct
equations (which were generally used numerically), there are no
changes to most constraints presented in other references.  The
only corrections apply to fits of energy-independent
suppressions; that is the fits no longer apply to the exclusion
of some regions of maximally mixed neutrinos.  Finally, we have
noted that the energy dependence of the day-night effect can be a
strong discriminator between various solutions of the solar
neutrino problem.

\acknowledgments{
We would like to thank Paul Schechter for a useful conversation. 
We also greatly appreciate Kristin Burgess, Yang Hui He, and Jun
Song for reviewing the paper.  MS would like to thank the
National Science Foundation (NSF) for her gracious fellowship,
and the Air Force Institute of Technology (AFIT) for supporting
this research. MS would like to thank Dr.\ Krastev for helping us
understand the parameters involved in calculating the contour
plots for the day-night effect. We would also like to thank
Robert Foot for useful comments on the manuscript. This work is
supported in part by funds provided by the U.S. Department of
Energy (D.O.E.) under cooperative research agreement
\#DF-FC02-94ER40818.}

\newpage
\appendix

\section{Derivation of MSW equations}

\label{MSWDerivation}

In this appendix we present a derivation of the MSW effect. First
we derive the MSW equations of motion for an individual neutrino. 
We then find the energy eigenstates of the system and use them to
find the wave function amplitudes for electron neutrinos produced
in the sun and evolved into the vacuum.  To describe the ensemble
of neutrinos we introduce the density matrix.  After averaging
out the rapid oscillations we find a steady state solution to the
density matrix equations of motion.  We average this solution
over the regions of neutrino production.

We begin by finding the MSW equations of motion for an individual
neutrino.  The coupling describing the interaction between
electron neutrinos and electrons is
\begin{equation}
H_{int} = \sqrt{2} G_F N_e,
\label{Hint}
\end{equation}
where $N_e$ is the number density of electrons. This contribution
to the interaction Hamiltonian is added to the Schr\"{o}dinger
equation written in the flavor basis.  We assume that
$|\nu_e\rangle$ can be written as a superposition of only two
mass eigenstates, $|\nu_1\rangle$ and $|\nu_2\rangle$.  We let
$|\nu_x \rangle$ denote the orthogonal linear combination of
$|\nu_1\rangle$ and $|\nu_2\rangle$, which might be any
superposition of $|\nu_\mu \rangle$ (muon neutrino) and
$|\nu_\tau\rangle$ (tau neutrino).  The transformation between
the $\nu_1$-$\nu_2$ and $\nu_e$-$\nu_x$ bases is then given by
\begin{equation}
\left( \matrix{ {C_{\nu_1}} \cr {C_{\nu_2}} \cr } \right) =
\left( \matrix{ \cos\theta_V & -\sin\theta_V \cr
\sin\theta_V & \cos\theta_V \cr } \right)
\left( \matrix{ {C_{\nu_e}} \cr {C_{\nu_x}} \cr } \right) \ ,
\end{equation}
where the variable $\theta_V$ is the vacuum mixing angle, and
$C_{\nu}\equiv \langle \nu | \Psi \rangle$ for $\nu =$ $ \nu_1$,
$\nu_2$, $\nu_e$ or $\nu_x$.  This equation can be written
compactly by introducing the index notation
\begin{equation}
C_{\nu_i}=U^\dagger_{if} C_{\nu_f} \ ,
\label{Uconvention}
\end{equation}
where the repeated index $f$ is summed over $\nu_e$ and $\nu_x$,
and $i$ is summed over the mass eigenstates. The Schr\"{o}dinger
equation for this system is:
\begin{equation}
i \partial_t \left( \matrix{ {C_{\nu_e}} \cr {C_{\nu_x}} \cr }
\right)= \left[
U \left( \matrix{ p+\frac{m_1^2}{2p} & 0 \cr 0 &
p+\frac{m_2^2}{2p} \cr } \right) U^\dagger + \left( \matrix{
{\sqrt{2}G_F N_e} & 0 \cr 0 & {0} \cr }\right)
\right]
\left( \matrix{ {C_{\nu_e}} \cr {C_{\nu_x}} \cr } \right),
\end{equation}
where we have expanded the energy in the ultra-relativistic limit
so that $E=p+\frac{m^2}{2p}$. We now substitute $U$ into the
Schr\"{o}dinger equation, obtaining
\begin{equation}
i \partial_t \left( \matrix{ {C_{\nu_e}} \cr {C_{\nu_x}} \cr }
\right)=
\left( \matrix{B & A \cr
     A & - B \cr } \right)
\left( \matrix{ {C_{\nu_e}} \cr {C_{\nu_x}} \cr } \right)
\label{eom}
\end{equation}
where
\begin{eqnarray}
\Delta_0 & \equiv & \frac{1}{2p}(m_2^2-m_1^2) \\
 A & \equiv & \frac{\Delta_0}{2} \sin2 \theta_V 
\label{AdefApp} \\
 B & \equiv & \frac{\sqrt{2}}{2} G_F N_e -\frac{\Delta_0}{2} \cos2
\theta_V ,
\label{BdefApp}
\end{eqnarray}
and where we have dropped the term $ p+
\frac{(m_1^2+m_2^2)}{4p} + \frac{\sqrt{2}}{2} G_F N_e$ which is
proportional to the identity, because terms proportional to the
identity cannot contribute to mixing.

The eigenvalues are $\pm \lambda(N_e)$, where $ \lambda(N_e)=
\sqrt{A^2 + B^2}$, and the eigenvectors are:
\begin{equation} v_- =\left( \matrix{ \sqrt{\frac{\lambda-B}{2 \lambda}} \cr
-\sqrt{\frac{\lambda+B}{2 \lambda}} \cr }
\right) {\rm \ ,\ and \ } v_+ =\left( \matrix{
\sqrt{\frac{\lambda+B}{2 \lambda}} \cr
\sqrt{\frac{\lambda-B}{2 \lambda}} \cr }
\right).\end{equation}
Since these eigenvectors form the matrix that will diagonalize
the interaction matrix in the presence of matter, it is useful to
parameterize them by a matter mixing angle $\theta_M(N_e)$:
\begin{equation} \cos\theta_M = \sqrt{\frac{\lambda-B}{2 \lambda}} \mbox{,\
\ \ and\ \ }
\sin\theta_M = \sqrt{\frac{\lambda+B}{2 \lambda}}, 
\label{mmangledef}
\end{equation}
or equivalently
\begin{equation} \lambda \cos 2 \theta_M = - B, \end{equation}
or
\begin{equation} \lambda \sin2 \theta_M = A .\end{equation}
Defining the matrix
\begin{equation} U(\theta_M)=\left( \matrix{\cos \theta_M & \sin \theta_M
\cr -\sin \theta_M & \cos \theta_M \cr
} \right) \ ,
\label{matterU} 
\end{equation}
the Hamiltonian can be diagonalized as
\begin{equation} 
U(\theta_M) \left( \matrix{- \lambda & 0 \cr 0 & \lambda \cr }
\right) U^\dagger(\theta_M) =
\left( \matrix{B
& A \cr A & - B \cr } \right) .
\end{equation} 
We maintain the notation introduced in Eq.~(\ref{Uconvention}) so
that $C_{\nu_i}(\theta_M)=U^\dagger_{if}(\theta_M) C_{\nu_f}$ in
or out of matter, where $C_{\nu_i}(\theta_M) \equiv \langle \nu_i
| \Psi \rangle$ denotes the amplitude for the overlap of the
neutrino state with instantaneous mass eigenstates $|\nu_i
\rangle$.

\label{AdiabaticDerivation}

To describe the evolution of the neutrinos as they travel to the
earth from their creation point in the sun, it is useful to
develop the adiabatic approximation, in which one assumes that
the density changes imperceptibly within an oscillation length. 
Remembering that $U$, $\theta_M$, and $\lambda$ are all functions
of the local electron density $N_e$, and hence functions of time,
we write the Schr\"{o}dinger equation in the basis
$\nu_i(\theta_M)$ of the instantaneous mass eigenstates:
\begin{eqnarray}
i \partial_t \left( \matrix{ {C_{\nu_1}} \cr {C_{\nu_2}} \cr }
\right) & = & \left( \matrix{- \lambda & 0 \cr 0 & \lambda \cr }
\right) \left( \matrix{ {C_{\nu_1}} \cr {C_{\nu_2}} \cr }
\right)+ (i\partial_t U^\dagger ) U \left( \matrix{ {C_{\nu_1}} \cr
{C_{\nu_2}} \cr } \right) \\
& = &
 \left( \matrix{- \lambda & 0 \cr 0 & \lambda \cr }\right)
\left( \matrix{ {C_{\nu_1}} \cr {C_{\nu_2}} \cr } \right)+
 i\left( \matrix{ 0 & - \partial_t \theta_M \cr
            \partial_t \theta_M & 0 \cr } \right)
  \left( \matrix{ {C_{\nu_1}} \cr {C_{\nu_2}} \cr } \right).
\end{eqnarray}
The adiabatic approximation is the assumption that the
off-diagonal terms $\partial_x \theta_M$ can be neglected, in
which case the equation is easily integrated:
\begin{equation}
\left( \matrix{ {C_{\nu_1}(t_f)} \cr {C_{\nu_2}(t_f)} \cr }
\right) \approx
\left( \matrix{ e^{+i \phi(t_f)} & 0 \cr 0 & e^{-i \phi(t_f)} \cr }
\right)
\left( \matrix{ {C_{\nu_1}(t_0)} \cr {C_{\nu_2}(t_0)} \cr }
\right) \ ,
\end{equation}
where
\begin{equation}
\phi(t_f) = \int_{t_0}^{t_f} \lambda(t) d t \ .
\end{equation}
Because the adiabatic states form a complete basis, we can always
write the exact solution as a superposition of the two adiabatic
states.  This final superposition is expressed by two unknown
variables, $a_1$ and $a_2$ where $|a_1|^2 + |a_2|^2 = 1$. The
$|a_2|^2$ parameter represents the probability of a non-adiabatic
transition, which is most likely to happen when the neutrinos
cross resonance, the density at which $B=0$, when the two
eigenvalues become nearly equal. Likewise $|a_1|^2=1$ would
represent adiabatic evolution.  Given any initial state
$\nu_f(t_0)$ in the flavor basis, the final state can be written
in the general form:
\begin{equation} 
\left( \matrix{ {C_{\nu_1}(t_f)} \cr {C_{\nu_2}(t_f)} \cr }
\right) =
\left( \matrix{a_1 & a_2 & \cr -a_2^* & a_1^* & \cr } \right)
\left( \matrix{ e^{+i \phi(t_f)} & 0 \cr 0 & e^{-i \phi(t_f)} \cr }
\right)
U^\dagger\Bigl(\theta_M(t_0)\Bigr)\ \nu_f(t_0) \ .
\end{equation}
For an electron neutrino originating in a medium of mixing angle
$\theta_M$, the above equation implies that the final state in
the vacuum is given by
\begin{equation}
\left( \matrix{ C_{\nu_1}(t_f) \cr C_{\nu_2}(t_f) \cr}\right)
\equiv  \left( \matrix{ A_1 \cr A_2 \cr} \right) =
\left( \matrix{ a_1 \cos\theta_M e^{+i\phi} + a_2
\sin\theta_M e^{-i\phi} \cr -a_2^* \cos\theta_M e^{+i\phi} +
a_1^* \sin\theta_M e^{-i\phi} \cr} \right) \ .
\label{FinalState}
\end{equation}

We now go on to talk about the ensemble of neutrinos reaching the
earth. To describe a quantum mechanical ensemble of neutrinos, it
is useful to introduce the density matrix
\begin{equation}
\rho \equiv \sum_i f_i | \nu_i \rangle \langle \nu_i | \ ,
\end{equation}
where $f_i$ denotes the probability that the particle is in the
quantum state $| \nu_i \rangle $.  The density matrix
corresponding to a single neutrino as described by
Eq.~(\ref{FinalState}) is therefore given by
\begin{equation}
\rho = \left( \matrix{ |A_1|^2 & A_1 A_2^* \cr A_1^* A_2 &
     |A_2|^2 \cr} \right) \ ,
\end{equation}
where
\begin{eqnarray}
|A_1|^2 & = & {1 \over 2} \left[ 1 + \cos 2 \theta_M \left( 1 - 2
|a_2|^2 \right) \right] + {1 \over 2} \left[ a_1 a_2^* \sin 2
\theta_M e^{2 i \phi(t_f)} + c.c \right] \\
A_1 A_2^* & = & {1 \over 2} \sin 2 \theta_M \left[ a_1^2 e^{2 i
\phi(t_f)} - a_2^2 e^{- 2 i \phi(t_f)} \right] - a_1 a_2 \cos 2
\theta_M \\
|A_2|^2 & = & {1 \over 2} \left[ 1 - \cos 2 \theta_M \left( 1 - 2
|a_2|^2 \right) \right] - {1 \over 2} \left[ a_1 a_2^* \sin 2
\theta_M e^{2 i \phi(t_f)} + c.c \right]
\end{eqnarray}
In Appendix \ref{DeCoherance} we explain why this process allows
us to eliminate the terms that have rapidly oscillating phases. 
In particular, the phase angle $\phi(t_f)$ and the phases of the
complex numbers $a_1$ and $a_2$ are all rapidly varying functions
of the neutrino energy, the location in the sun where the
neutrino is produced, and the precise time of day and year at
which the neutrino is observed. The density matrix which
describes the ensemble of observed neutrinos is constructed by
averaging over these quantities, so any quantity with a rapidly
oscillating phase will average to zero.  This is equivalent to
the statement that the $\nu_1$ and $\nu_2$ components arriving at
the earth are incoherent, so we average over their phases.  The
matrix elements of the phase-averaged density matrix are given by
\begin{eqnarray}
\langle |A_1|^2 \rangle & = & {1 \over 2} \left[ 1 + \cos 2
\theta_M \left( 1 - 2 |a_2|^2 \right) \right] \\
\langle A_1 A_2^* \rangle & = & 0 \\
\langle |A_2|^2 \rangle & = & {1 \over 2} \left[ 1 - \cos 2
\theta_M \left( 1 - 2 |a_2|^2 \right) \right] \ .
\end{eqnarray}

The term $| a_2 |^2 \equiv P_{{\rm jump}}$ is the probability of
crossing from one adiabatic state to the other during the time
evolution of these operators.  An approximate expression for
$P_{{\rm jump}}$ can be found by using a linear approximation for
the density profile at resonance \cite{Parke}, yielding
\begin{equation}
P_{{\rm jump}}= \exp \left( -\frac{\pi \Delta m^2 \sin^2(2
\theta_V) N(x_{{\rm res}})} {4 p \cos (2 \theta_V)
N'(x_{{\rm res}})} \right).
\end{equation}
Here $N(x_{{\rm res}})$ is the density at the point where the
neutrino crosses resonance, and $N'(x_{{\rm res}})$ is the first
derivative of the density at resonance. More accurate
approximations to $P_{\rm{jump}}$ and the details of their
derivation can be found in Refs.~\cite{hepph9712304,hepph9606353}
and the references therein.

The density matrix corresponding the ensemble of observed
neutrinos must be obtained by averaging over the production sites
in the sun.  While we have already made use of this fact in
dropping all terms with rapidly oscillating phases, we must still
average the slowly varying terms which remain.  Letting $^8B(r)$
denote the normalized probability distribution for production at
a distance $r$ from the center of the sun, one finds
\begin{equation}
\rho = \left( \matrix{ {1 \over 2} + C_0 & 0 \cr 0 & {1 \over 2} - C_0 \cr}
\right) \ ,
\end{equation}
where
\begin{equation}
C_0={1 \over 2} \int_0^{R_{{\rm sun}}} dr\ ^8B(r)\
\cos\Bigl(2\theta_M(r)\Bigr)\ (1-2 P_{{\rm jump}}) \ .
\end{equation}
Note that the diagonal entries of $\rho$ are just the fractions
$k_1$ and $k_2$ of $\nu_1$ and $\nu_2$ flux from the sun,
respectively, as defined in Sec.~\ref{derivation}.  Therefore
\begin{equation}
k_1 = {1 \over 2} + C_0 \ , \ k_2 = {1 \over 2} - C_0 \ .
\end{equation}

Finally, we transform to the $\nu_e$-$\nu_x$ basis, so
\begin{equation} 
\rho^{\nu_e \nu_x} = U(\theta_V) \rho U^\dagger
(\theta_V) \equiv \left( \matrix{\rho_{ee} & \rho_{ex} \cr
\rho_{xe} & \rho_{xx}\cr} \right) \ .
\end{equation}
One then finds that the probability of observing a neutrino
reaching the surface of the earth as an electron neutrino is
given by
\begin{equation}
P_S= \rho_{ee} = {1 \over 2} + C_0 \cos2\theta_V \ .
\label{rhoee}
\end{equation}
The off-diagonal matrix element is given by
\begin{equation}
\rho_{x e} = - C_0 \sin2\theta_V \ .
\label{rhoem}
\end{equation}

Our numerical simulations were all performed by integrating
Eq.~(\ref{eom}) to solve for $P_{2e}$, and also by integrating
the density matrix equations of motion. The evolution of the
density matrix is given by
\begin{equation}
i \hbar \partial_t \rho = - [ \rho , H ] .
\label{dm_evolution}
\end{equation}
Using Eq.~(\ref{dm_evolution}) with the Hamiltonian in the flavor
basis, we find that our new equations of motion are
\begin{equation} i \partial_t \rho_{ee} = A (\rho_{x e} - \rho_{x e}^*) \end{equation}
\begin{equation} i \partial_t \rho_{x e} = 2 (A \rho_{ee} - B \rho_{x e} ) - A\ ,
\end{equation}
where $A$ and $B$ are defined in Eqs.~(\ref{AdefApp}) and
(\ref{BdefApp}).  This allows us to perform calculations using
the complete mixed ensemble.  The expressions given in
Eqs.~(\ref{rhoee}) and (\ref{rhoem}) form a steady state solution
of the density matrix equations of motion in the vacuum. 
For typical ($\Delta m^2 \approx 1\times 10^{-5} \ {\rm eV}^2$
and $E=8 \ {\rm MeV}$) maximally mixed Boron-8 ($^8B$) neutrinos,
$P_{2e}$ oscillates and can take any value between $0$ and $1$. 
$C_0 \approx -{1 \over 2}$, which means that $k_2
\approx 1$, and Eq.~(\ref{Psehalf}) reduces to $P_{SE} = P_{2e}$. 
Thus $P_{SE}$ exhibits oscillatory behavior, and is in no way
constrained to be $1/2$ at maximal mixing. 
\section{Calculation methodology for the day-night effect}
\label{CalcDetails}
\begin{figure}
\centerline{\psfig{figure=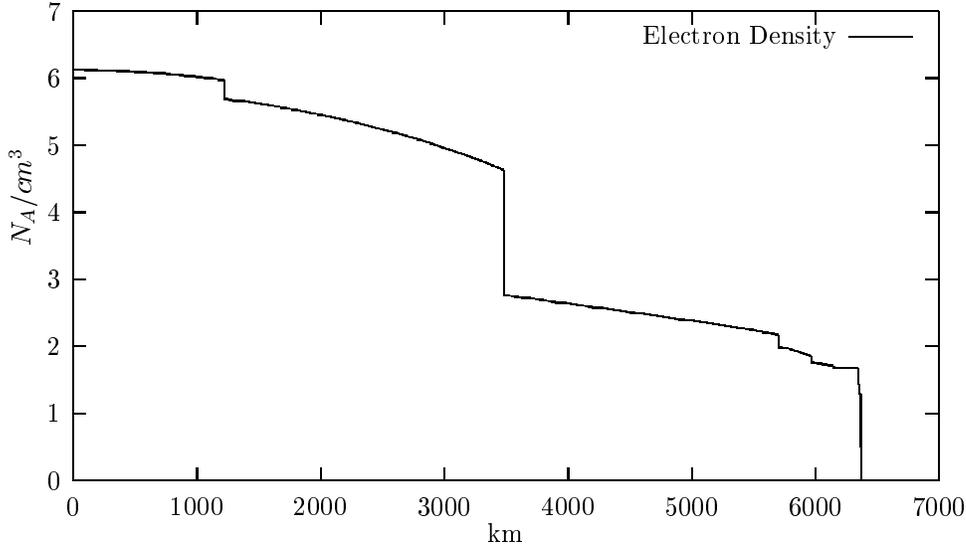,angle=0}}
\caption{The Preliminary Reference Earth Model (PREM) electron
density ($N_e$) profile of the earth.  $N_e$ is shown in units of
Avogadro's number of electrons per cm$^3$.}
\label{EarthProfile}
\end{figure}
First we calculated $P_S$ using Eq.~(\ref{rhoee}) for the
spectrum of $\Delta m^2 / p$ at various mixing angles. For a
given $\Delta m^2 / p$ we averaged $P_S$ over the regions of
$^8B$ neutrino production in the sun, provided by
Ref.~\cite{WebBahcall}.  Using $P_S$ to describe the neutrinos
that arrive at the earth, we then performed the evolution through
the earth with the density matrix equations of motion.  The
initial conditions for the density matrix are given by
\begin{equation}
\rho_{ee}= P_S,
\end{equation}
and
\begin{equation}
\rho_{xe}= -{1 \over 2} (2 P_S -1 ) \tan2 \theta_V.
\end{equation}
At maximal mixing we assume that $\rho_{xe}=-{1 \over 2} \cos\Bigl(2
\theta_M(t_0)\Bigr) \sin2\theta_V \approx {1 \over 2}$ which is the
adiabatic result.  This assumption is justified because in the
regions of parameter space under consideration near maximal
mixing, $P_{{\rm jump}}\approx 0$.  It follows that in these same
regions of parameter space the evolution remains adiabatic in the
limit where $\theta_V = \pi/4$. We use the earth density profile
given in the Preliminary Reference Earth Model (PREM) \cite{PREM}
(see Fig \ref{EarthProfile}).  To convert from the mass density
to electron number density we use the charge to nucleon ratio
$Z/A = 0.497$ for the mantle and $Z/A=0.467$ for the core.  The
numerical calculations were performed using a fourth order
Runge-Kutta integration programmed in C++. We propagated the
neutrinos through the earth for 90 zenith angles, $\alpha$,
evenly spaced between 90 and 180 degrees. We calculate the
anticipated electron flux as a function of zenith angle and
energy, denoted $P_{SE}(\alpha,E_\nu)$. The calculation
parameters are chosen for those of the Super-Kamiokande detector.
The normalized $^8B$ neutrino spectrum, $\Phi(E_\nu)$, and solar
electron densities, $N_e$, are also obtained from data-files
provided by Ref.~\cite{WebBahcall}.  Effective neutrino cross
sections are available which take into account the electron
recoil cross section with radiative corrections, the energy
resolution, and the trigger efficiency
\cite{electronspectra,WebBahcall}. We used these more accurate
cross sections for the overall day-night effect plotted in
Fig.~\ref{ContourDM}. Because these effective cross sections
already include the integration over detected electron recoil
energy, to calculate the recoil electron spectrum we used the
differential neutrino-electron scattering cross sections given in
Ref.~\cite{nuAstrophysics}. Using these data files and numerical
results the cross section for the scattering of solar neutrinos
of energy $E_\nu$ with electrons to produce a recoil electron
of energy $T'$ at the zenith angle $\alpha$ is given by
\begin{equation}
\frac{d \sigma_{\nu_{\rm solar}}}{dT'}(T',E_\nu,\alpha) =
P_{SE}(\alpha,E_\nu) \frac{d \sigma_{\nu_e}}{dT'}(T',E_\nu) +
\left[ 1-P_{SE}(\alpha,E_\nu)\right] \frac{d
\sigma_{\nu_\mu}}{dT'}(T',E_\nu) .
\end{equation}
Since muon and tau neutrinos have the same neutral current
interactions, we can use the $\nu_\mu$ cross section for the
$\nu_x$. The analysis of the recoil electron spectra is
explained in Refs.~\cite{electronspectra}. The actual flux at
recoil energy, $T$, is
\begin{equation}
g(\alpha,T)= \int_0^\infty dE_\nu\ \Phi(E_\nu)
\int_0^{T'_{{\rm max}}} dT'\ R(T,T')\ \frac{d \sigma_{\nu_{\rm
solar}}}{dT'}(T',E_\nu,\alpha)
\end{equation}
where the energy resolution of the detector is incorporated
through
\begin{equation}
R(T,T')=\frac{1}{\Delta_{T'} \sqrt{2 \pi}}
\exp\left(\frac{-(T'-T)^2}{2\Delta^2_{T'}}\right).
\end{equation}
The energy resolution, $\Delta_{T'}$, around the true electron
energy $T'$ for Super-Kamiokande is given by
\begin{equation}
\Delta_{T'}=(1.6\ {\rm MeV})\ \sqrt{T'/(10\ {\rm MeV})}.
\end{equation}
To calculate the average day-night effect over one year, we
weight the flux by the zenith angle exposure function $Y(\alpha)$
explained in Appendix \ref{zdf}.  The daytime measured flux at a
given measured electron recoil energy, $T$, is given by
\begin{equation}
D(T)=\int_0^{90} d \alpha\ g(\alpha,T) Y(\alpha),
\end{equation}
and for nighttime is
\begin{equation}
N(T)=\int_{90}^{180} d \alpha\ g(\alpha,T) Y(\alpha).
\end{equation}
The day-night asymmetry as a function of recoil electron energy
plotted in Fig.~\ref{Adnvsnrg} is given by
\begin{equation}
A_{d-n}(T)=\frac{N(T)-D(T)}{N(T)+D(T)}.
\end{equation} 
The final day-night asymmetry plotted in Fig.~\ref{ContourDM} is
given by
\begin{equation}
A_{d-n}= \frac{ \int_{5\ {\rm MeV}}^\infty dT\ \Bigl( N(T)-D(T)
\Bigr)}{ \int_{5\ {\rm MeV}}^\infty dT\ \Bigl( N(T)+D(T) \Bigr)}
\end{equation}
where 5 MeV is the minimum energy detected at Super-Kamiokande.
Verification of the accuracy of the computer code has been
accomplished with the help of \cite{Krastev}, and by comparing
our simulations to plots and data available in the literature.
\section{Validity of the steady state approximation}
\label{DeCoherance}
\begin{figure}
\centerline{\psfig{figure=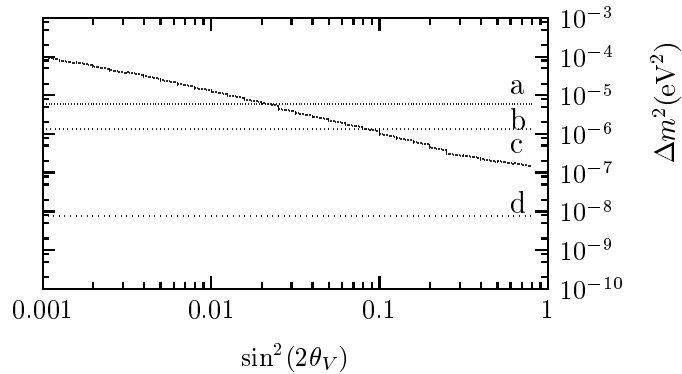,angle=0}}
\caption{The regions satisfying the conditions for steady state
density matrix. Above the line (a) is in steady state because of
wave packet separation.  Above the line (b) can be treated as
steady state because of the eccentricity of the earth's orbit. 
Above the diagonal line (c) is in steady state because the region
producing the neutrinos is much larger than an oscillation
length, and this phase averaging survives until the neutrinos
reach the vacuum.  Above line (d) is in steady steady state
because of the energy resolution of our detectors.}
\label{ExclusionPlot}
\end{figure}

Most of the work in the past decade on the MSW effect has assumed
that the ensemble of neutrinos reach the earth in a steady state
solution of the density matrix (i.e., in an incoherent mixture of
the mass eigenstates $\nu_1$ and $\nu_2$). There are several
reasons that the neutrinos reach the earth in a steady state: (a)
The separation of the $| \nu_1 \rangle $ and $| \nu_2 \rangle $
wave-packets while propagating from the sun to the earth exceeds
the size of the individual wave packets, eliminating the
interference effects.  (b) The eccentricity of the earth's orbit
results in a daily change of the earth-sun radius larger than the
vacuum oscillation length of the neutrinos.  (c) The neutrinos
are produced in a region much larger than their local oscillation
length.  (d) The energy resolution of the current detectors
coupled with the earth-sun radius perform an average. We now
proceed to map out parameter space justifying where the steady
state approximation is valid.
First we consider the separation of the two eigenstates during
transit to the earth.  This results in system that is an
incoherent superposition of $|\nu_1\rangle$ and $|\nu_2\rangle$.
	The width of the wave-packets, $\sigma_x$, is given by 
	Ref.~\cite{Kim}: 
\begin{equation} \sigma_x \approx 0.9 \times 10^{-7} \ {\rm cm}. \end{equation}
This results in a coherence length given by:
\begin{equation}
L_{coh}=2 \sqrt{2} \sigma_x \frac{ 2 E^2}{\Delta m^2}.
\label{CohLength}
\end{equation}
We lose coherence between the mass eigenstates if $L_{coh} < 1\
{\rm AU} = 1.5 \times 10^{13}\ {\rm cm}$.  If we require that the
incoherence condition apply up to 14 {\rm MeV} to include all
$^8B$ neutrinos, we find that for all of $\sin^22 \theta_V$ where
$\Delta m^2\ >\ 6.63 \times 10^{-6} \ {\rm eV}^2$ the
wave-packets have separated upon reaching earth. This corresponds
to the region labeled (a) in Fig.~\ref{ExclusionPlot}. Because
there is a continuous beam of neutrinos arriving from the sun, we
can ignore the fact that the lighter mass eigenstate arrives
first, and simply drop terms that rapidly oscillate due to the
lack of interference between the two states.

In the previous case the interference effects vanish because of a
loss of coherence between the mass eigenstates for a neutrino
produced at a specific place and time. In the remaining topics
the interference effects vanish due to averaging over the
ensemble of neutrinos which reach the detector.  Regardless
        of the source of the incoherence, any two systems with
	the same density matrix behave identically in single counting 
	experiments \cite{hepph9802387}. 

Next, we analyze the effect of the eccentricity of the earth's
orbit .  We are interested in day-night effects; therefore, if
the earth-sun radius changes more than an oscillation length
during one day, this will result in washing out any phase
dependence in the results measured over a period of one year. 
Between perihelion and aphelion the earth-sun radius changes $2 e
(1\ {\rm AU}) = 5.1 \times 10^{11}\ {\rm cm}$, where $ e=0.017$
is the earth eccentricity.  The earth-sun radius changes by this
quantity once each 180 days giving an average daily change in
radius of $2.83\times 10^{9}${\rm cm}.  This ensures our
incoherent phase for $ \Delta m^2 \ >\ 1.2 \times 10^{-6}\ {\rm
eV}^2 $.  This region is denoted by everything above the line
marked (b) in Fig.~\ref{ExclusionPlot}.

Third, we study the impact of where the neutrinos were produced. 
If the neutrino region of production is greater than the local
oscillation length of the neutrinos, then neutrinos of all
possible phases exist in the ensemble.  For a continuous beam of
neutrinos, this also results in dropping the rapidly oscillating
terms.  The condition is satisfied for the entire parameter space
under consideration $0.001 < \sin^2 2 \theta_V < 1$ and $1\times
10^{-11} \ {\rm eV}^2 < \Delta m^2 < 1\times 10^{-3} \ {\rm
eV}^2$.  However, one must be careful in making this statement. 
Although the region of production may be greater than the
neutrino oscillation length in the sun, the neutrinos could
undergo a non-adiabatic transition bringing a specific phase into
dominance.  This is the case for vacuum oscillations ($\Delta m^2
\approx 4 \times 10^{-10} \ {\rm eV}^2 $). The $^8B$ neutrinos
are produced mostly at $R_{^8B}=0.046\ R_{sun}=3.2\times 10^9{\rm
cm}$.  The vacuum oscillation length is on the order of 1
\hbox{AU}. However the oscillations length near the solar core
where these neutrinos are produced is about $1.8\times 10^{7}{\rm
cm} \ll R_{^8B}$. Although the neutrinos are produced in a region
larger than their oscillation length, they acquire roughly the
same phase in the process of leaving the sun. This occurs because
the density change upon leaving the sun occurs more rapidly than
the oscillation length of the neutrinos, violating the condition
of adiabaticity. To express this quantitatively we estimate that
if $P_{{\rm jump}} < 0.1 $ for $14 \ {\rm MeV}$ neutrinos that
the initial randomly distributed oscillation phases at the time
of production will persist as the neutrinos leave the sun and
enter the vacuum. This leads to a steady state solution
applicable in the parameter space above the diagonal line labeled
(c) shown in Fig.~\ref{ExclusionPlot}.

Last, we study the impact of the energy resolution on our ability
to discriminate phases. Assuming perfect coherence between the
two mass eigenstates the phase upon reaching the earth is given
by
\begin{equation}
\phi = \frac{\Delta m^2 (1 \ {\rm AU})}{4 p \hbar c}.
\end{equation}
Our uncertainty in energy impacts our uncertainty in phase
through error propagation:
\begin{equation}
\delta \phi= \left| \frac{d\phi}{dp}\right| \delta p =
\frac{\Delta m^2 (1 \ {\rm AU})}{4 p^2 \hbar c} \delta p.
\end{equation}
If our uncertainty in our phase is greater than $2 \pi$ we are
again justified in treating our ensemble as a steady state. Using
conservative figures for energy ($p=14\ {\rm MeV}$), and the
energy resolution ($\delta p \approx 1\ {\rm MeV}$)
\cite{electronspectra}, we find that for $\Delta m^2 > 6.5 \times
10^{-9} \ {\rm eV}^2$ we are justified in the steady state
approximation. This corresponds to parameter space above the line
labeled (d) in Fig.~\ref{ExclusionPlot}. Recently
Ref.~\cite{hepph9903329} also reached the same conclusions
outlined in this appendix.

\section{The zenith distribution function}
\label{zdf}
\begin{figure}
\centerline{\psfig{figure=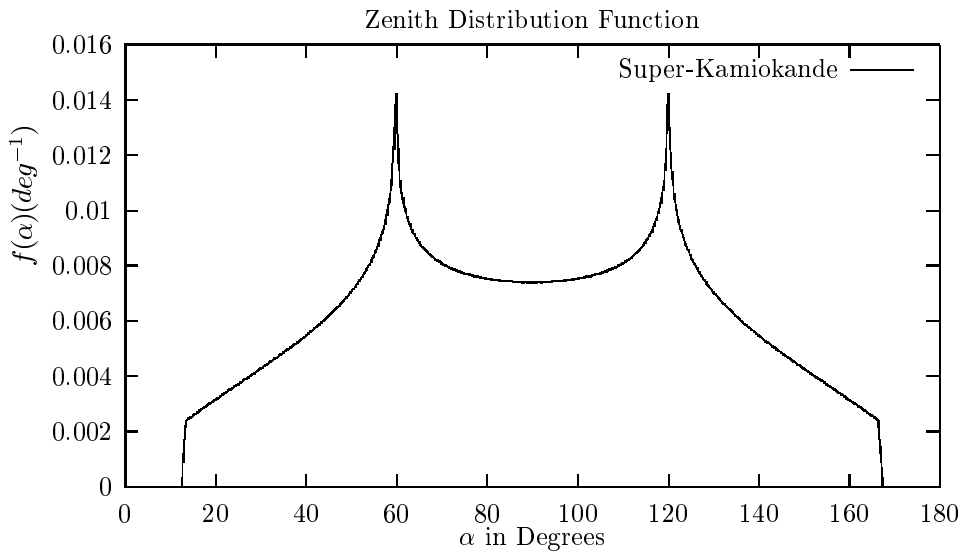,angle=0}}
\caption{The zenith distribution function at Super-Kamiokande.}
\label{ZDF}
\end{figure} 
The zenith angle distribution function gives the fraction of the
time that the sun is at a given zenith angle. The function is
calculated by numerically simulating the orbit of the earth
around the sun.  We begin by writing the vector towards the
zenith of the detector in coordinates for which the earth's orbit
lies in the $x$-$y$ plane:
\begin{equation}
\vec{r}_z =
\left( \matrix{ 1 & 0 & 0 \cr 0 & \cos\delta & -\sin\delta
\cr 0 & \sin\delta &
   \cos\delta \cr } \right)
 \left(\matrix{  \sin(90^\circ-L) \cos \phi  \cr
            \sin(90^\circ-L) \sin \phi  \cr
            \cos(90^\circ-L)  \cr } \right)
\end{equation}
where the north latitude is given by $L$, $\phi$ gives the time
of day in radians, and $\delta = 23.439^\circ$ is the earth's
declination \cite{AA96}. Because we are averaging over a one year
time period we can arbitrarily choose the initial time of year,
and the initial time of day. The vector pointing from the earth
towards the sun is
\begin{equation}
\vec{r}_s=\left(\matrix{  \cos D   \cr
 \sin D  \cr 0 \cr} \right)
\end{equation} 
where $D$ is the day of the year in radians. From here we can
find the local zenith angle from the dot product $\vec{r}_s \cdot
\vec{r}_z =\cos \alpha $. To numerically calculate the zenith
function distribution we divided $\alpha$ into 360 bins evenly
spaced between $0$ and $\pi$. Now we run $ 0 \leq D \leq 2 \pi$
and $0 \leq \phi \leq 2 \pi$ over 1000 steps in $D$ and 1000
steps in $\phi$ and count how much relative time $\alpha$ spends
over each bin.  We generate the zenith angle distribution
function for Super-Kamiokande which sits in Gifu Prefecture,
Japan at $36.43^\circ$ north latitude \cite{hepex9812009}.  This
produces the undistorted zenith function distribution seen in
Fig.~\ref{ZDF}.  One can also obtain this function as a data file
from \cite{WebBahcall} which includes small corrections for the
eccentricity of the earth's orbit and the wobble of the earth's
declination.  To maximize accuracy we performed our calculations
using this data file.



\bibliographystyle{JHEP.bst}
\providecommand{\href}[2]{#2}\begingroup\raggedright\endgroup

\end{document}